# Baselining Wireless Internet Service Development - An Experience Report

Fabio Bella, Jürgen Münch, and Alexis Ocampo

**Abstract** — New, emerging domains such as the engineering of wireless Internet services are characterized by a lack of experience based on quantitative data. Systematic tracking and observation of representative pilot projects can be seen as one means to capture experience, get valuable insight into a new domain, and build initial baselines. This helps to improve the planning of real development projects in business units. This article describes an approach to capture software development experience for the wireless Internet services domain by conducting and observing a series of case studies in the field. Initial baselines concerning effort distribution from the development of two wireless Internet pilot services are presented. Furthermore, major domain-specific risk factors are discussed based on the results of project retrospectives conducted with the developers of the services.

## 1 Introduction

The engineering of wireless Internet services is an emerging application domain characterized by quickly evolving technology, upcoming new devices, new communication protocols, support for new, different types of media, and varying and limited communication bandwidth, together with the need for new business models that will fit in with completely new service portfolios. Examples of new wireless Internet services can be expected in the domains of mobile entertainment, telemedicine, travel services, tracking and monitoring services, or mobile trading services.

Due to its recentness, this domain lacks explicit experience related to technologies, techniques, and suitable software development process models that is based on quantitative data. Unreliable project planning, incorrect effort estimates, and high risk with respect to process, resource, and technology planning, as well as with regard to the quality of the resulting product are inevitable consequences of this lack of experience. One means to capture experience and get valuable insight into a new domain is systematic tracking and observation of representative pilot projects.

This paper presents a study consisting of two case studies aimed at quantitative baselining. Additionally, the case studies were used to gain qualitative experience. The article aims at giving managers and developers a sense of the behavior of projects in the wireless Internet domain.

The approach followed in this study is based on a combination of descriptive process modeling, GQM-based measurement, and collection of lessons learned: Descriptive process modeling is applied in order to understand and improve the software development process as applied within the observed organizations; GQM-based measurement is practiced to gather quantitative experience, whereas qualitative aspects are addressed by the retrospective-based collection of lessons learned. Therefore, this study should be seen as a challenging attempt to characterize a promising new application domain not only from a qualitative, but also from a quantitative point of view. Of particular interest is the focus placed on first effort distribution baselines gathered from the development of suitable pilot services.

Section 2 introduces the methodologies applied within the study and explains how they relate to it. Section 3 discusses the context in which the case studies were performed; the overall approach applied to gather quantitative as well as qualitative experience, and the results in terms of effort distribution baselines and major domain-specific risks observed. Section 4 subsumes the article and sketches future work to be performed.

## 2 Background

This study is based on a combination of descriptive process modeling [4], GQM-based measurement [5], [15], and retrospective-based collection of lessons learned [7]. This Section gives an overview of the methodologies applied and explains how they relate to the study.

The main idea of the descriptive process modeling is to explicitly document the development processes as they are applied within a given organization: A so-called process engineer observes, describes, and analyzes the software development process and its related activities, and provides descriptions of the processes to the process performers. Since the processes are usually complex, support is needed for both process engineers and process performers. Descriptive process modeling is applied within the context of the study with the help of the Spearmint® environment. The architecture of Spearmint® and its features for a flexible definition of views, used for retrieving filtered and tai-

- *F. Bella is with the Fraunhofer Institute for Experimental Software Engineering (IESE), Sauerwiesen 6, 67661, Kaiserslautern, Germany. E-mail: bella@iese.fraunhofer.de.*
- *J. Münch is with the Fraunhofer Institute for Experimental Software Engineering (IESE), Sauerwiesen 6, 67661, Kaiserslautern, Germany. E-mail: muench@iese.fraunhofer.de.*
- *A. Ocampo is with the Fraunhofer Institute for Experimental Software Engineering (IESE), Sauerwiesen 6, 67661, Kaiserslautern, Germany. E-mail: ocampo@iese.fraunhofer.de.*

lored presentations of process models, is presented in [4]. One distinct Web-based view, namely the Electronic Process Guide (EPG), is used for disseminating process information and guiding process performers, e.g., project managers and developers.

The Goal/Question/Metric (GQM) approach is applied to define measurement goals and a proper measurement infrastructure [5], [15]. During the first two steps, business and improvement goals are analyzed and metrics defined according to the process model elicited through Spearmint®The results of this first phase are GQM plans that comprise all metrics defined.

In the following step, the project plan and the process model are used to determine by whom, when, and how data are to be collected according to the metrics. The data collection procedures are the results of this instrumentation.

Raw data are collected according to the data collection procedures. The collected raw data are analyzed and interpreted according to the GQM plan and the feedback provided by the interested parties.

In the next step, the interested parties draw consequences based on the analysis and their interpretations.

Finally, analysis, interpretations, and consequences are resumed in the measurement results and collected as experience in the experience database for future reuse.

In addition to the measurement of quantitative data, the collection of qualitative data is driven by project retrospectives [7]. Therefore, meetings and interviews with the participants of different work packages are conducted regularly to focus lessons learned and improvement potentials. Concerning more specific wireless-related topics, published experience in the field were important sources of information, particularly at the beginning of the study. An extensive overview of related work is given by [9].

## 3 Characterizing Effort in the Wireless Internet Services Engineering Domain

In the following, the context of the study, the approach applied, and the main related results are described.

### 3.1 Context of the Case Studies

The present study was conceived as an integral part of the evaluation of the Wireless Internet Service Engineering (WISE) project. The project produces integrated methods and components (COTS and open source) to engineer services on the wireless Internet. The production of methods and components is driven by the development of pilot services.

The methods already produced include a reference architecture, a reference development process model, as well as guidelines for handling heterogeneous mobile devices.

The components include a service management component and an agent-based negotiation component.

Three pilot services, i.e., a financial information service, a multi-player game, and a data management service, are being developed by different organizations. The data from the development processes of two of the pilot services are the basis of this study.

The duration of the project is 30 months and an iterative, incremental development style is applied: three iterations are performed, of roughly 9 months each.

In iteration 1, a first version of the planned pilot services was built using GPRS. At the same time, a first version of methods and tools was developed.

In iteration 2, a richer second version of the pilots was developed on GPRS, using the first version of methods and tools. In parallel, an improved second version of methods and tools was developed.

In iteration 3, the final version of the pilots is being developed on UMTS, using methods and tools from the second iteration. Also, a final version of methods and tools is being developed.

Currently, the third iteration is still running. The case studies discussed in this article refer to data from the first two iterations.

### 3.2 Case Studies Method

This subsection presents the process-centric approach applied within the WISE project for gathering experience in the new domain.

As mentioned in the previous subsection, in parallel to the development of the pilot services, a measurement infrastructure was defined in order to evaluate the effects of the method and tools applied to develop these services. The infrastructure is based not only on measures but also on interviews and any other available evidence.

Figure 1 sketches the strategy applied iteratively during each of the three iterations to gather, package, and maintain experience from the development of the pilot services. The experience acquisition process is depicted using the Spearmint® notation: The circles represent activities, the rectangles artifacts, the arrows indicate produces/consumes relationships between activities and products.

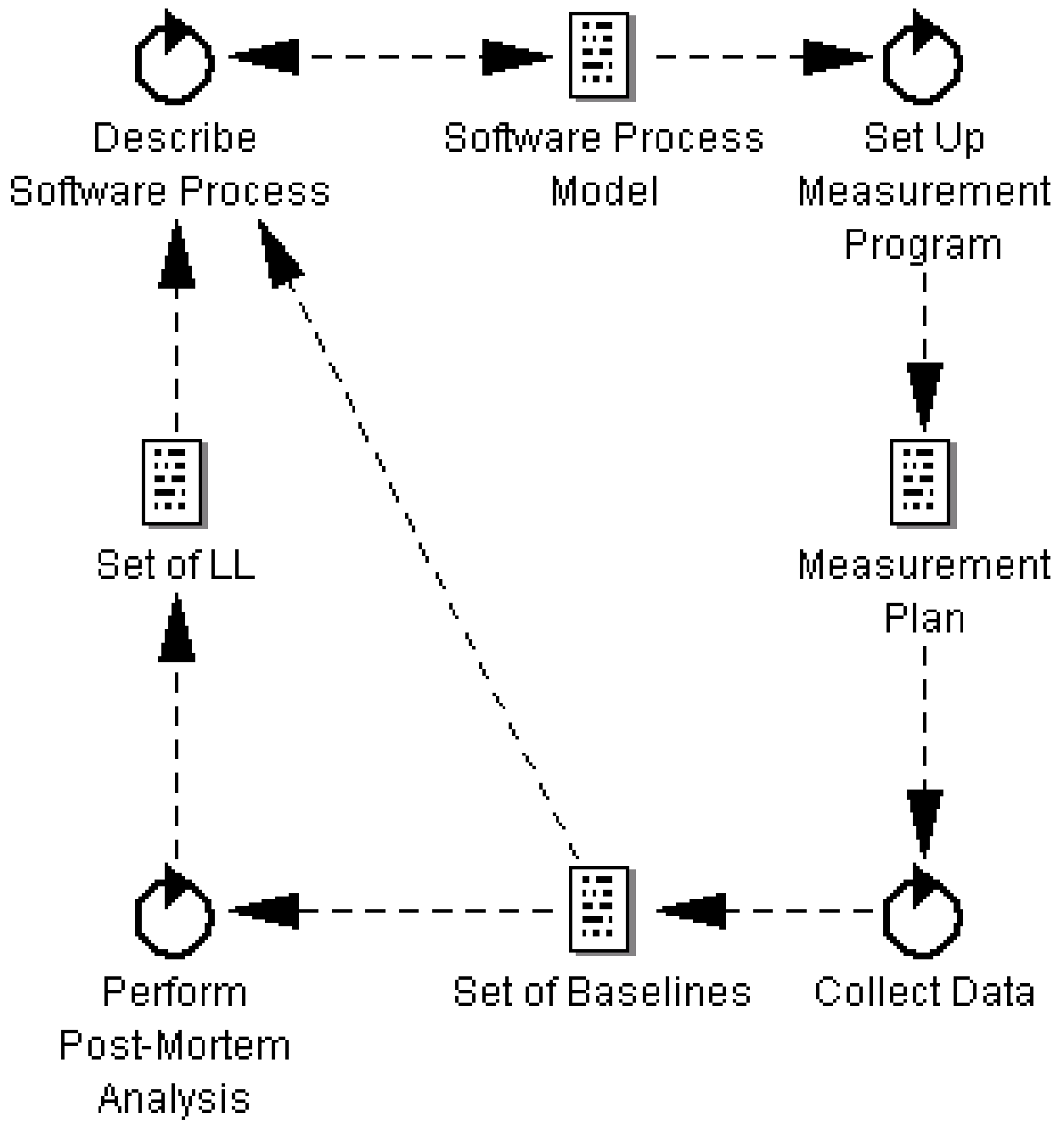

Fig. 1. Experience acquisition

At the beginning of each iteration, software development processes are elicited as applied by the organizations; the descriptive process models (in Figure 1, *Software Process Model*) are used to set up effort measurement programs (*Measurement Plan*).

During the development of the pilot services, the pilot performers collect data according to the measurement plans. The data is validated and stored. At the end of the development cycle, baselines are built, i.e., the data collected are aggregated and quality models are built (*Set of Baselines*).

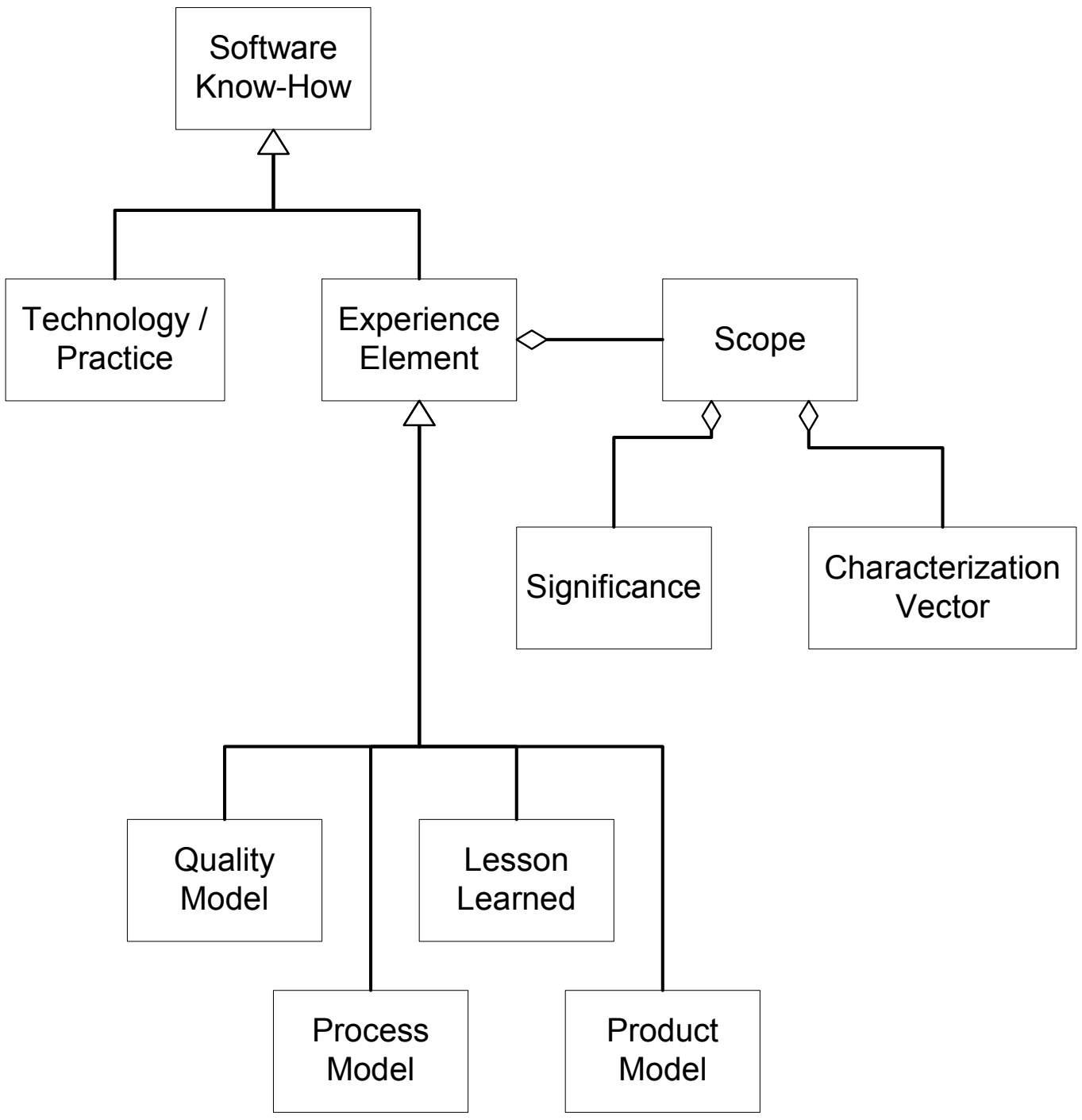

Fig. 2. Overview of experience elements

During post-mortem analysis sessions the baselines are discussed with the involved parties, then interpretations and consequences for the next iteration are worked out (e.g., the possible evolution of the surrounding development process). In order to get more insights of a qualitative nature, lessons learned (*Set of LL*) are collected regularly by interviewing project participants at project meetings or by phone. Many lessons were also gathered through the analysis and interpretation of baselines. Therefore, within the context of the WISE project, different experience models were applied (see Figure 2), which are an adaptation of basic principles of the Experience Factory [1] and QIP [2] approaches.

All kinds of software engineering experience are regarded as experience elements: process and product models, quantitative quality models (i.e., baselines), and qualitative experience (such as lessons learned). For each experience element, the scope of its validity is described.

The scope consists of a characterization vector and the significance. The characterization vector characterizes the environment in which the experience element is valid, i.e., the context surrounding a given project (see Table 1). The significance describes how the experience element has been validated and to which extent (e.g., validation through formal experiments, single case study, or survey).

TABLE 1
EXCERPT OF A CHARACTERIZATION VECTOR

| **Customization factor** | **Characteristic** | **Pilot X** |
|---|---|---|
| Domain characteristics | Application type | Computation-intensive system |
| | Business area | Mobile online entertainment services |
| Development characteristics | Project type | Client New development<br>Server New development |
| | Transport protocol | GSM/GPRS/UMTS |
| | Implementation language | Client: J2ME<br>Server: J2EE |
| | Role | Technology provider, service developer |

## 3.3 Results

This subsection presents the results gathered from the first two development iterations. The discussion of results focuses upon effort baselines from the development of the pilot services and major domain-specific risks observed.

### *3.3.1 Effort Baselines related to the Development of the Pilot Services*

This subsection discusses quality models concerning effort distribution. The quality models are gathered from the development of two pilot services.

**Case Study 1**

Context: Pilot service 1 provides a solution for real time stock tracking on mobile devices: the user can view real time quotes concerning a whole market or define his/her own watch lists. The partner responsible for this development is a provider of high end trading services on the Internet, aimed at banks and brokers. The pilot is the adaptation of an existing Web-based information service. Critical usability issues arise due to the huge amount of data needed by a financial operator to perform an analysis and the small-sized display of mobile devices. Furthermore, since the Internet traffic on mobile devices is paid for by the end user, based on data volume and not on connection time, and since frequent refresh of a large amount of financial data is required, the adoption of the push technology instead of the pull technology is an important issue, because it avoids unnecessary data refreshes for the user. Most of the usability issues were addressed during the first iteration. The second iteration was mainly concerned with implementing a solution based on the push technology.

The life cycle model applied for developing the pilot service during each iteration is an iterative process model consisting of three phases: a requirements phase, a development / coding phase, and a testing phase. The ad-hoc process is characterized by extensive use of verbal communication within the development team, and little use of ex-plicit documentation. Another important characteristic of the development process is the absence of an explicit design phase. This can be seen as a consequence of the fact that the overall system architecture and the related interfaces were known at the beginning of the project, since this was mainly the same client server architecture used to provide the service on the traditional Internet. The client side was a prototype developed using the Wireless Markup Language (WML); during the second iteration, the client was developed using the Java 2 platform, Micro Edition (J2ME). In both cases, the prototype and its high-level design were documented after development.

Analysis: The analysis of the effort distribution observed during the first iteration and represented in Figure 3 shows that most of the effort (approx. 84%) was spent on the development phase, i.e., the creation of the first prototype. Only approx. 15% of the overall effort was spent on the requirements phase. This can be explained as follows: the functional requirements were described at a high degree of abstraction, which was possible since they were derived from the available Internet service and they were therefore well understood; the more challenging non-functional requirements, e.g., usability issues, were not formalized at all, since they were not understood at the beginning of the project and they were to be investigated with the WML prototype.

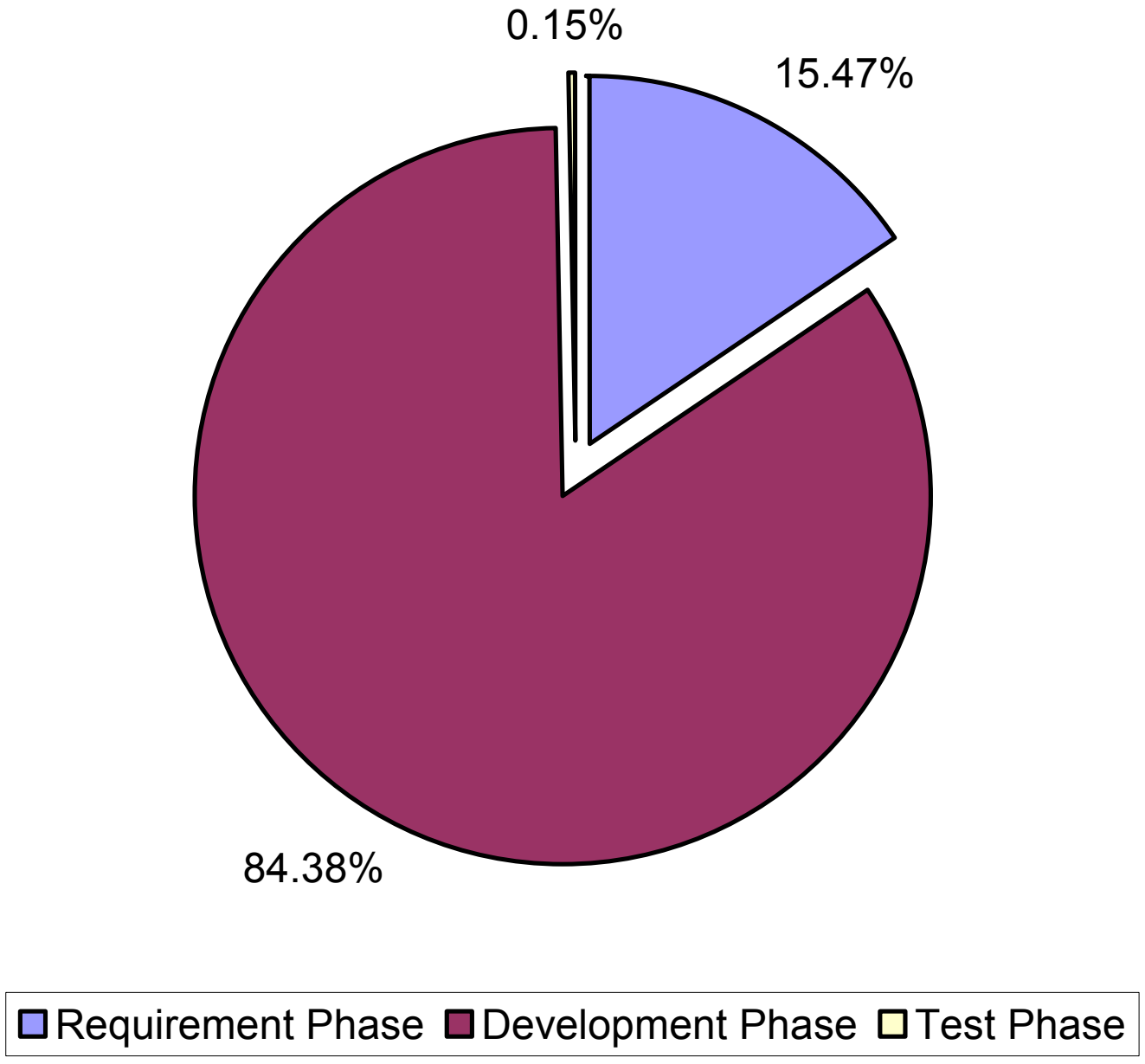

Fig. 3. Effort distribution, pilot service 1, iteration 1

Since more effort than planned was spent on the development phase, little effort remained to be spent on the testing phase.

During the first iteration, the development of pilot service 1 required about 340 man-days.

Figure 4 shows the effort distribution observed during the second iteration. The consequences of a more accurate description of the development process and, at the same time, of the stabilization of the process enacted by the development team became visible and, therefore, a different, more balanced, effort distribution can be observed. The greater amount of effort collected in the requirements phase can be attributed to a change of the underlying process model description. During the first iteration, it was noticed that a part of the effort collected in the development phase was spent on performing some feasibility studies rather than on implementing the prototype. The goal of the studies was to evaluate different mobile devices and WML constructs with respect to usability requirements. Therefore, for the second iteration, it was decided to collect the effort related to the feasibility studies as requirements phase.

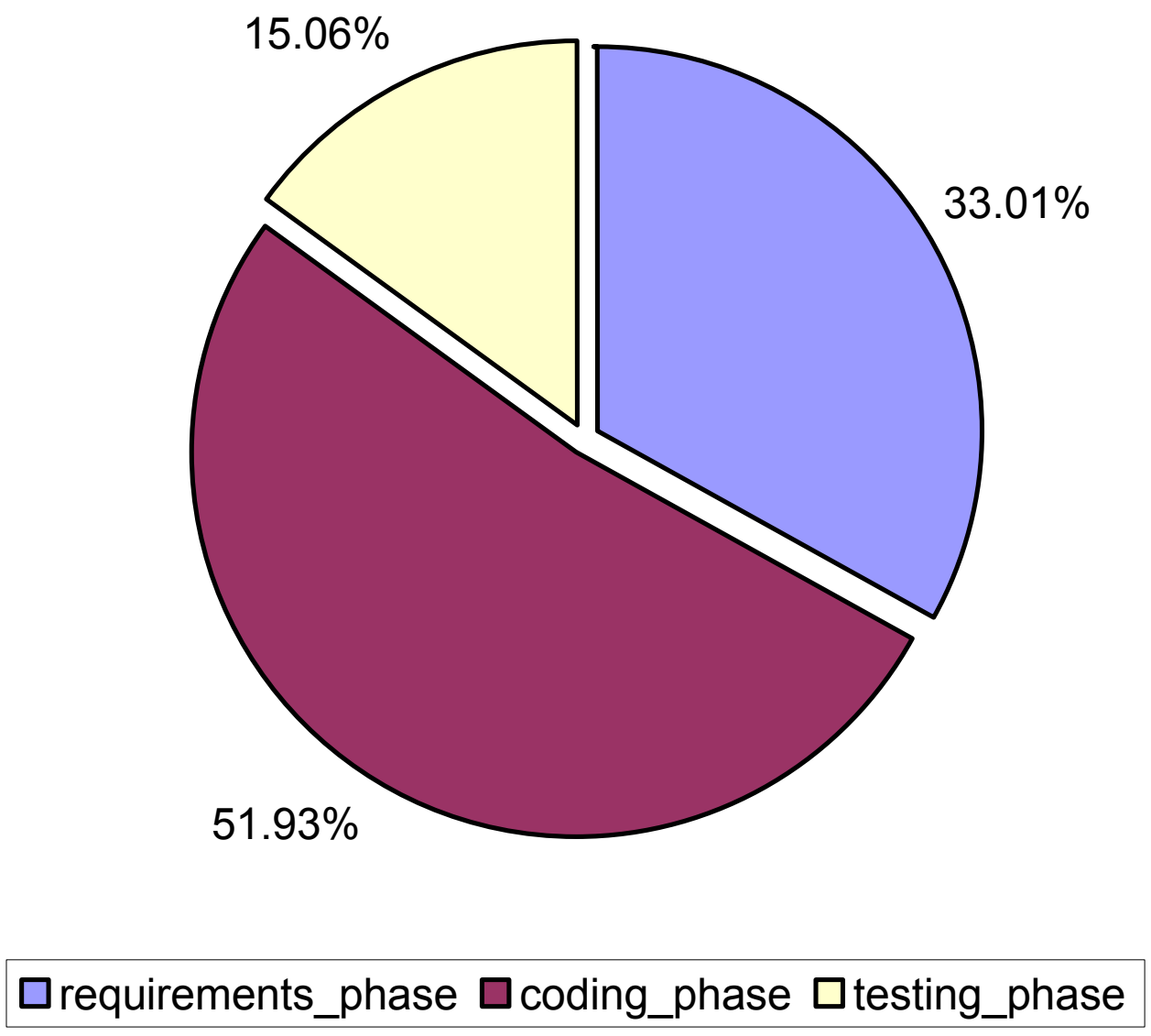

Fig. 4. Effort distribution, pilot service 1, iteration 2

Although more effort was spent on integration testing than during the first iteration, most of the effort collected as testing phase was spent on documenting the integrated code. Due to the fact that the coding phase was underestimated, most of the system test was shifted to the third iteration.

During the second iteration, the development of pilot service 1 required about 200 man-days.

**Case Study 2**

Context: Pilot service 2 is concerned with the new development of a multi-player online game for mobile devices: many users interact in a shared environment, i.e., a virtual labyrinth. The players can collect different items, chat, and fight against enemies and against each other. From a business point of view, games and entertainment could be, after voice and SMS, the next killer application on the wireless Internet. The development is distributed between two different teams / organizations: one organization is responsible for the development of the client on the mobile device and provides a multimedia-messaging stack on the terminal part; the other organization customizes the multimedia layer on the server side.

The organization responsible for the client side reaches CMM maturity level 3. An iterative life cycle model consisting of four phases (requirements phase, design phase, coding phase, and testing phase) was followed in this case within the context of each single iteration. The process is characterized by extensive use of verbal communication as well as of explicit formal documentation.

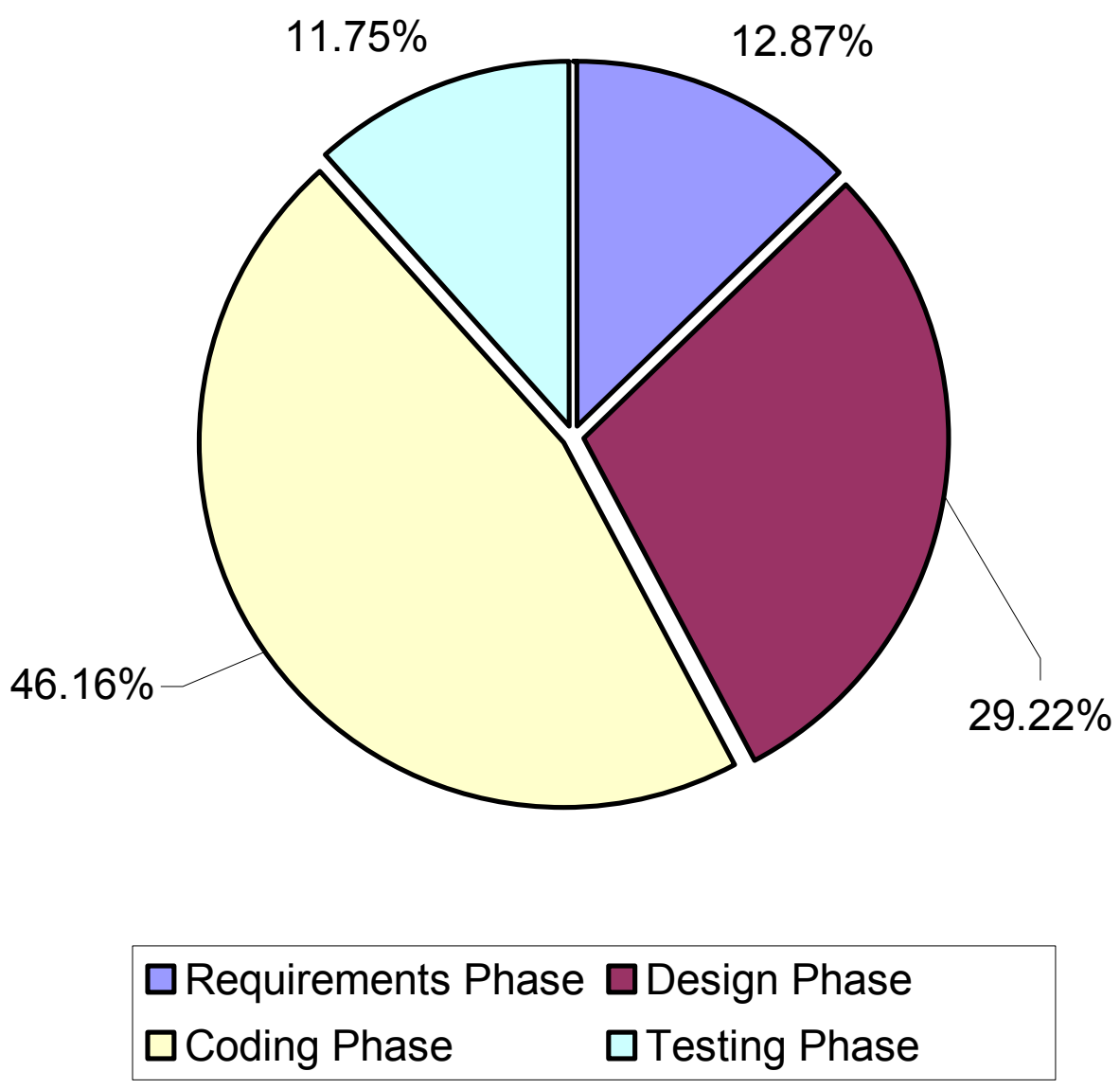

Fig. 5. Effort distribution, pilot service 2 (client side), iteration 1

Analysis (client side): During the first iteration, the following effort distribution was observed (see Figure 5): approx. 13% of the development effort was spent on the requirements phase, 29% on design, 46% on coding, and 12% on testing.

Unexpected problems were reported in the requirements and the design phase: problems in determining which organization should develop the server side led to unexpected low effort spent on the definition of the requirements; problems with the use of TCP/IP as transport protocol led to unexpected great effort in designing an alternative protocol on the basis of UDP. The problematic behavior of the TCP/IP protocol represents a good example of unexpected issues that may occur when applying common Internet technologies within the wireless context.

Finally, it was reported that less effort than planned was spent on testing.

During the first iteration, the development of the client side of pilot service 2 required about 140 man-days.

As depicted in Figure 6, during the second iteration, approx. 28% of the development effort was spent on the requirements phase, 15% on design, 50.5% on coding, and 6.5% on testing. In this case, too, unexpected problems were reported during the requirements phase, since the organization in charge of developing the server side left the project. On the other hand, due to a redesign of the graphic library that led to simplification of the further design, less effort than planned had to be spent on the design phase. It was also reported that, due to organizational issues, less effort than planned was spent on testing. As a consequence, an extensive system test must be performed during the third iteration.

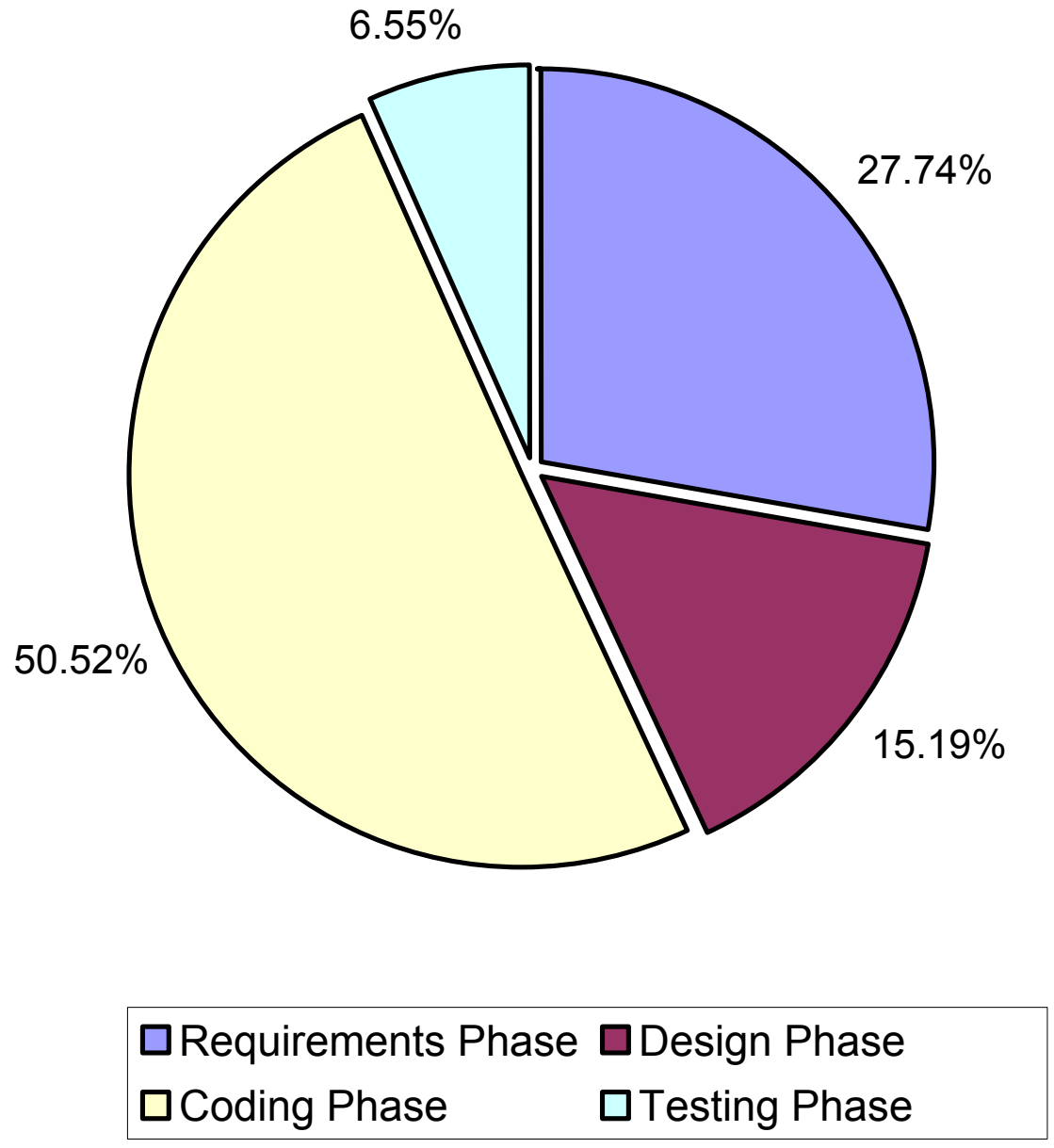

Fig. 6. Effort distribution, pilot service 2 (client side), iteration 2

During the second iteration, the development of the client side of pilot service 2 required about 130 man-days.

It should be noted that effort estimates were provided at the beginning of each iteration. In order to obtain more accurate estimates for the second iteration, the effort distribution data from the first iteration were used together with the first estimates as basis for the estimation process. Figure 7 shows how the new values for the new estimates were chosen from within a range between the data estimated before the beginning of the first iteration and the data gathered during the first iteration.

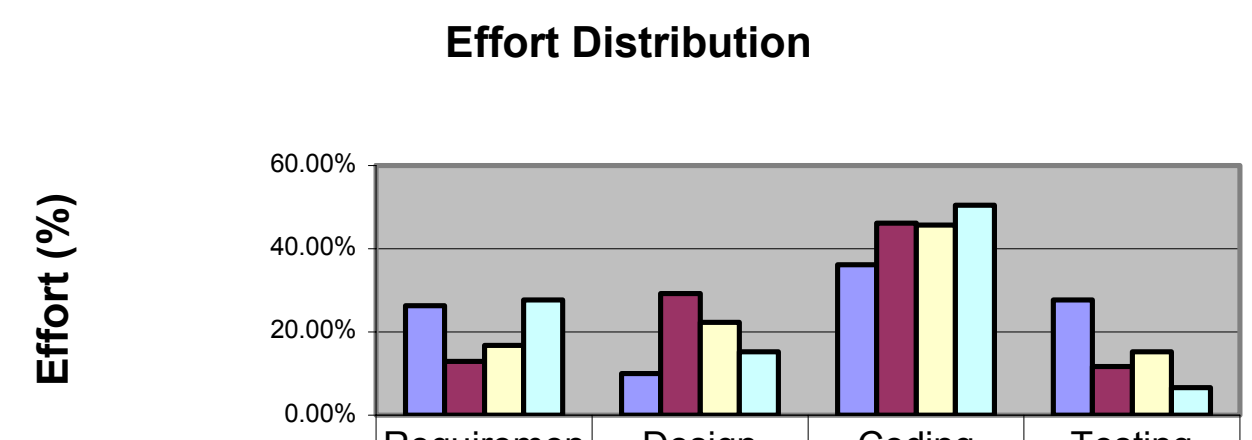

| | Requirements Phase | Design Phase | Coding Phase | Testing Phase |
|---|---|---|---|---|
| Estimate It. 1 | 26.31% | 9.98% | 36.09% | 27.62% |
| Effort It. 1 | 12.87% | 29.22% | 46.16% | 11.75% |
| Estimate It. 2 | 16.74% | 22.36% | 45.69% | 15.21% |
| Effort It. 2 | 27.74% | 15.19% | 50.52% | 6.55% |

Phases

Fig. 7. Overview of effort distributions and effort distribution estimates during the first two iterations

Concerning the requirements phase, for example, approx. 26% was the estimate for the first iteration, 13% was the effort actually spent on this phase during the first iteration, and 17% was estimated for the second iteration. The new estimated value is less than the estimate from the first iteration, but greater than the value actually measured. The estimation values were also chosen according to the critical issues expected in the second iteration.

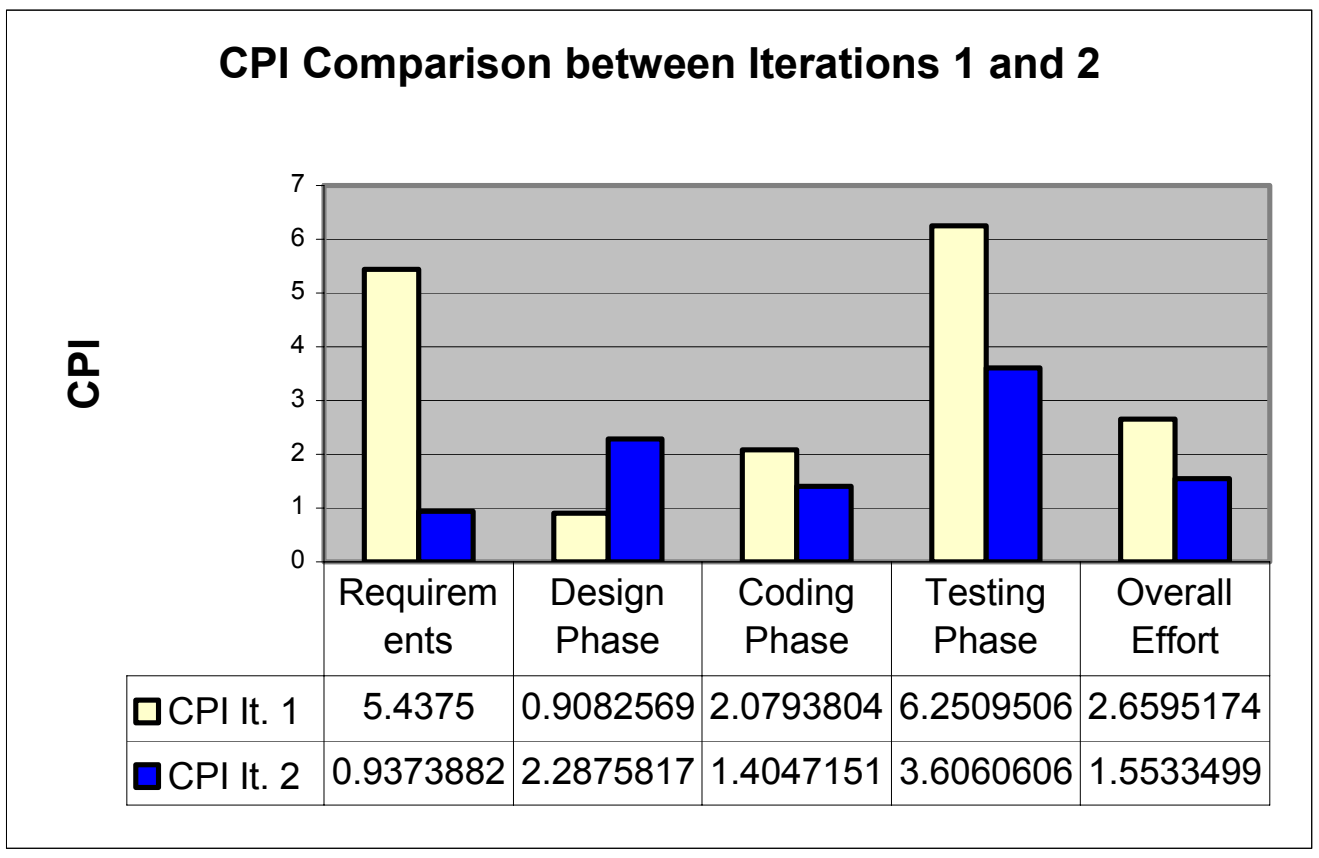

| | Requirements | Design Phase | Coding Phase | Testing Phase | Overall Effort |
|---|---|---|---|---|---|
| CPI It. 1 | 5.4375 | 0.9082569 | 2.0793804 | 6.2509506 | 2.6595174 |
| CPI It. 2 | 0.9373882 | 2.2875817 | 1.4047151 | 3.6060606 | 1.5533499 |

Fig. 8. Comparison of the cost performance indices from the first two development iterations of pilot service 2 (client side)

The comparison of the Cost Performance Indices (CPI = planned effort / actual effort [8]) computed during the two iterations and represented by Figure 8 shows that the effort estimates for the second iteration were more accurate than the estimates for the first iteration (according to the definition of CPI, an estimate is very accurate for CPI values close to 1, like the estimate concerning the requirements phase of the second iteration; values greater than 1 indicate overestimation, as in the case of the requirements phase during the first iteration).

Furthermore, during the first iteration, much additional effort was spent on management-related activities, like configuration management, project planning / tracking, and project support. Due to this, the effort spent on these activities was measured during the second iteration: it was seen that approx. 82% of the overall effort (1007.5 hours) was spent on development in the strict sense, whereas 18% (223 hours) was spent on management-related activities.

Analysis (server side): As mentioned above, two different organizations were in charge of developing the server part of pilot service 2. During the second iteration, the second organization extended the system developed during the first iteration.

Due to organizational issues, the requirements were managed by the organization responsible for the client side. As a consequence, both organizations in charge of the server side spent little effort on defining the requirements.

During the first iteration, an iterative life cycle model was adopted. As shown in Figure 9, effort was spent on design (32.5%), coding (52%), and integrating the client with the server part (15.5%). No requirements phase and no acceptance test were performed. Unexpected problems were reported during the design phase, which were caused by the TCP/IP protocol, whose latency was too high when used on GPRS.

During the first iteration, the development of the pilot service 2 server side required about 130 man-days.

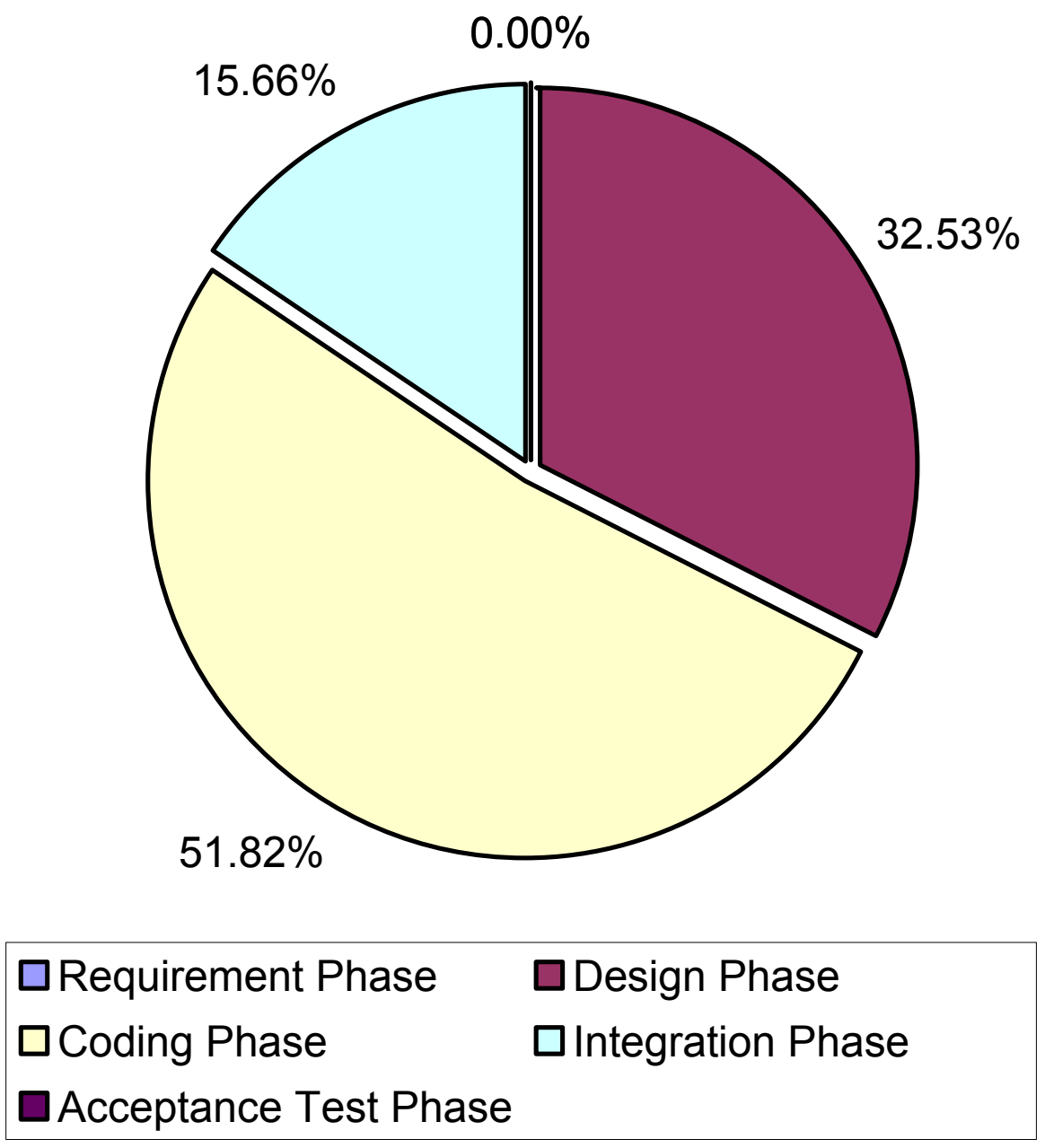

Fig. 9. Effort distribution, pilot service 2 (server side), iteration 1

During the second iteration, the organization involved in the development of the server side tried to apply an approach based on extreme programming. This makes it difficult to compare the effort data from the first and second iteration of the server part.

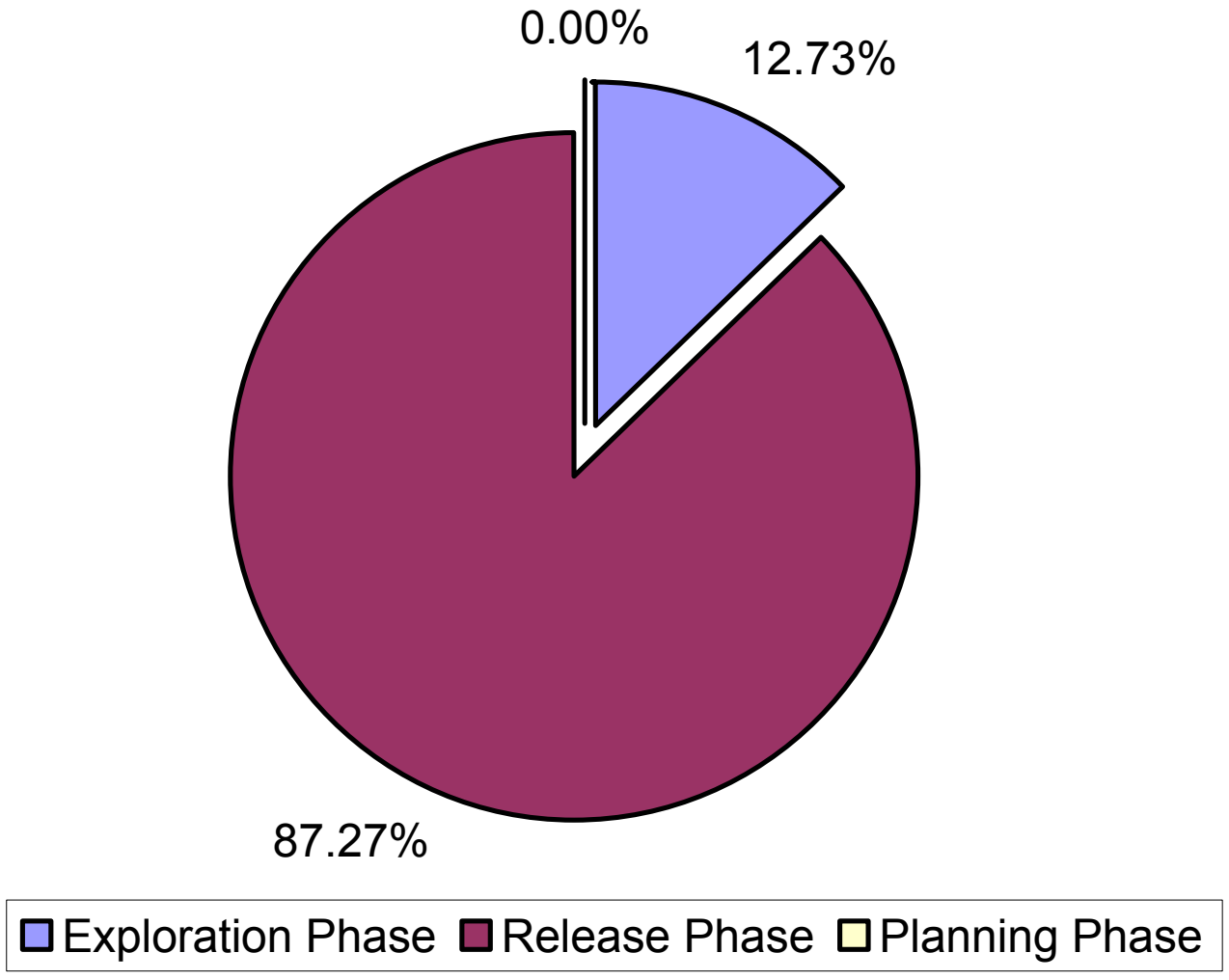

Fig. 10. Effort distribution, pilot service 2 (server side), iteration 2

Furthermore, for various reasons, extreme programming was not followed strictly. Figure 10 shows, for example,

that the planning phase was not performed, since all requirements and their related priorities had already been defined by the organization responsible for the client side. Also, many difficulties were encountered in deploying the server developed by the first organization in the new organization's environment; these facts did not allow many short development cycles and related releases as foreseen by the XP approach. As a consequence, the whole development was performed at once in one big cycle. Moreover, the lack of experience with the test first technique led to unexpected effort and, although the technique was recognized to be very interesting, it had to be given up.

During the second iteration, the development of the server side of pilot service 2 required about 80 man-days.

**Comparative Analysis**

In both case studies, the requirements phase was difficult to control due to the novelty of the domain and the fact that low level requirements and, particularly, usability-related requirements (e.g., how to represent large tables on small displays) were often not well understood at the beginning. Feasibility studies introduced in the second iteration proved to be a good means to make explicit and handle the related uncertainty.

This uncertainty is one of the reasons why the effort spent on the design phase was in all observed cases less than the effort spent on coding (max. 33% in the case of the development of the server side of pilot service 2 during the first iteration).

The effort data from the development of the pilot services showed that all organizations spent most of the development effort on coding (46% - 84% of development effort). This seems to be plausible if the many open issues are considered that could be addressed only at the coding level.

Testing proved very challenging due the great diversity of devices available on the market, the unreliability of device specifications, the low degree of automation of the testing procedures on real devices, and the unreliability of the available emulators. The effort spent until the end of the second iteration is considered by all the involved organizations to be insufficient, with the consequence that most of testing will be performed during the last iteration.

### 3.3.2 Domain-specific Risks

During the first two iterations, qualitative experience was collected by interviewing people involved in the development of the pilot services. Due to the novelty of the domain, the pilot partners had to deal with several risks that were unknown or at least not well understood at the beginning of the project. In the following, the main domain-specific risks observed during the development of the services are presented.

**R1**: The first issue to be considered is the great diversity of target devices in terms of display size and mode (i.e., resolution and number of colors), memory capacity, processor performance, and interaction mechanisms with the user (i.e., keyboard, jog dial, cursor buttons, joystick, touch screen, voice control, etc.). This heterogeneity makes it very difficult to reconcile the need for portability with the increasing demand for appealing applications.

**R2**: Java's promise of code working on every platform is difficult to achieve: different levels of compliance with the J2ME specification in the case of virtual machines implemented by different device manufacturers can lead to great variations in performance and behavior of the same application running on different mobile devices.

**R3**: The maturity of the technologies specific to the wireless domain should be carefully considered: many quality aspects of mobile devices (file system, network access capabilities, memory, etc.) are of a much lower level than those of regular desktop systems. This has consequences in terms of predictability of the quality of services and the development process.

**R4**: Technologies proven to be reliable when applied within the context of the traditional Internet may turn out to be unreliable or perform poorly when used within the context of the wireless world.

**R5**: Testing wireless Internet services proved very challenging due to different reasons: The first reason are the many usability issues (e.g., consistent interfaces, navigation, access, etc.) related to the great diversity of devices available on the market. Most of the usability issues have to be further researched due to the novelty of the domain. Another reason is the development for future announced devices: Device specifications are subject to change without notice and are usually unreliable. Another reason is that a lot of effort has to be spent on setting a proper environment. Emulators represent one unsatisfactory but necessary alternative solution. The main advantage of using emulators is the automation of the testing procedures whereas unreliable behavior is their greatest disadvantage.

### 3.3.3 Limits of the Study

Concerning the validity of the quantitative part of the study, i.e., the characterization of the effort distribution, Spearmint® EPGs played a major role in assuring consistent views on the different development processes. These views and the GQM approach were very helpful in defining sound measurement programs that proved suitable to provide correct and meaningful data on a monthly basis. The training of the developers responsible for collecting data was challenging due to the widely distributed project environment and some personnel changes that occurred between the two iterations.

Concerning the comparability of the quantitative data, it is not possible to directly compare either the numerical data from the different pilots or all data from different iterations. This is due to the different surrounding processes applied to develop the pilot services and the evolution of the processes during the whole project life cycle. Moreover, despite an extensive literature search, no studies could be found with a similar focus on effort baselines.

Concerning the generality of the results, the context of the single case studies defines the scope of validity of the baselines presented. Transforming the results for similar contexts should be done with careful analysis of the external validity.

Referring to the validity of the qualitative part of the study, i.e., the collection of lessons learned, the roles played within the pilots by the persons interviewed (mainly developers) and the focus of the respective pilots influenced the lessons reported.

Additionally, the domain-specific risks presented in this study should be regarded as being of high significance, since they were generalized from the lessons learned provided by the individual organizations involved in the development of the pilots.

## 4 Conclusions

This study aimed at providing effort baselines for managers and developers in order to give them a sense of the behavior of projects in the field of wireless Internet service engineering. Of course, it is important to mention that each project is different, and that the context in which the pilots were developed must be taken into consideration before making any type of analogies.

The effort data from the development of the pilot services showed that all organizations spent most of the development effort on coding. As expected, the requirements phase was characterized by a great degree of uncertainty concerning performance and availability of related technologies as well as many usability issues related to the great heterogeneity of the devices on the market.

Testing proved very challenging due the great diversity of devices available on the market, the unreliability of device specifications, the low degree of automation of the testing procedures on real devices, and the unreliability of the available emulators. As a consequence, defect characterization is a difficult task, and a great amount of the effort planned for the third iteration will be spent on it. It is still unclear how to characterize defects concerning usability issues. For this purpose, usability reports will be introduced in the next iteration.

The descriptive process modeling approach supported by the Spearmint® environment played a key role in stabilizing the processes, eliciting accurate process models, and disseminating process information to the process performers. These are all necessary preconditions for meaningful effort tracking and planning.

As expected, and in spite of the accurate process models, effort estimation proved to be a challenging process at the beginning. During the first iteration, the organizations involved were not able to deliver effort estimates or the estimates they delivered turned out to be inaccurate at the end of the iteration. On the other hand, effort tracking performed during the first iteration together with estimation processes based on the effort data collected provided more accurate effort estimates for the second iteration.

How to characterize complexity and/or size of system is still an open issue. Metrics for complexity/size can be useful for deriving effort estimation. Also, defect density measures can be built on them for controlling the testing process. In any case, in order to estimate effort on the basis of estimates of system size or complexity, much more research should be done. For example, regarding a metric like the number of lines of code (LOC), in the case of code written for mobile devices, it was seen that the number should be reduced in order to improve performance; also, a low number of classes is often the result of a great optimization effort and not necessarily evidence of a simpler module with less features.

## Acknowledgment

We would like to thank the WISE consortium, especially the pilot partners, for their fruitful cooperation. We would also like to thank Sonnhild Namingha from the Fraunhofer Institute for Experimental Software Engineering (IESE) and Jussi Ronkainen from VTT Electronics for reviewing the first version of the article.

## References

[1] Basili, V.R., Caldiera, G., Rombach H.D.: The Experience Factory, in Encyclopedia of Software Engineering (John J. Marciniak, Ed.), John Wiley & Sons, Inc., Vol. 1, pp. 469-476 (1994).

[2] Basili, V.R, Quantitative Evaluation of Software Engineering Methodology, in Proceedings of the First Pan-Pacific Computer Conference, Melbourne, Australia (1985).

[3] Becker-Kornstaedt, U., Boggio, D., Muench, J., Ocampo, A., Palladino, G.: Empirically Driven Design of Software Development Processes for Wireless Internet Services. Proceedings of the Fourth International Conference on Product-Focused Software Processes Improvement (PROFES) (2002).

[4] [4] Becker-Kornstaedt, U., Hamann, D., Kempkens, R., Rösch, P., Verlage, M., Webby, R., Zettel, J.: Support for the Process Engineer: The Spearmint Approach to Software Process Definition and Process Guidance. Proceedings of the Eleventh Conference on Advanced Information Systems Engineering (CAISE '99), pp. 119-133. Lecture Notes in Computer Science, Springer-Verlag. Berlin Heidelberg New York (1999).

[5] [5] Briand, L.C., Differding, C., Rombach, H.D: Practical Guidelines for Measurement-Based Process Improvement. Software Process Improvement and Practice 2, No.4, pages 253-280 (1996).

[6] Münch, J.: Muster-basierte Erstellung von Software-Projektplänen, PhD Theses in Experimental Software Engineering, Vol. 10, ISBN: 3-8167-6207-7, Fraunhofer IRB Verlag (2002).

[7] Kerth, N.L.: Project Retrospectives: A Handbook for Team Reviews. Dorset House Publishing, ISBN: 0-932633-44-7, New York (2001).

[8] Humphrey, W. S.: A Discipline for Software Engineering (SEI Series in Software Engineering). Carnegie Mellon University, ISBN: 0-201-54610-8. Addison-Wesley Publishing Company (1995).

[9] Ocampo, A.; Boggio, D.; Münch, J.; Palladino, G.: Toward a Reference Process for Developing Wireless Internet Services. In: IEEE Transactions on Software Engineering 29 (2003), 12, 1122-1134 : Ill., Lit.

[10] Jedlitschka, A.; Nick, M.: Software Engineering Knowledge Repositories. In: Conradi, Reidar (Ed.) u.a.: Empirical Methods and Studies in Software Engineering : Experiences from ESERNET. Berlin : Springer-Verlag, 2003, 55-80: Ill., Lit. (Lecture Notes in Computer Science 2765).

[11] Cockburn, A.: Agile Software Development: Addison-Wesley Pub. Co; ISBN: 0201699699; 1st edition (2001).

[12] Beck, K.: Extreme Programming Explained: Embrace Change. Addison Wesley (2000).

[13] Boehm, B.W.: A Spiral Model for Software Development and Enhancement, IEEE Computer, vol 21, No 5, pp. 61-72 (1988).

[14] Buchanan, G., Farrant, S., Jones, M., Thimbleby, H., Marsden, G., Pazzani, M.J.: Improving Mobile Internet Usability. In Proceedings World Wide Web 10, pp. 673-680 (2001).

[15] Solingen, R. van; Berghout, E.: The Goal/ Question/ Metric Method. A Practical Guide for Quality Improvement of Software Development. London, McGraw-Hill, 1999.

**F. Bella** received his MS in Computer Science from the Technical University of Kaiserslautern, Germany, in 2002. MS Thesis: “Design and Implementation of a Similarity Analysis between Process Models in Spearmint/EPG”. He developed his MS thesis at the Fraunhofer Institute for Experimental Software Engineering (IESE) in Kaiserslautern. Since October 2002, Fabio Bella has been a research scientist at IESE in the department of Quality and Process Engineering. Since June 2003, he is a Provisional SPICE Assessor. Bella’s research interests in software engineering include: (1) modeling and measurement of software processes and resulting products, (2) software quality assurance and control, (3) technology transfer methods, and (4) software process assessments.

**J. Münch** received his PhD degree (Dr. rer. nat.) in Computer Science from the University of Kaiserslautern, Germany. Dr. Münch is Department Head and Competence Manager for Quality and Process Engineering at the Fraunhofer Institute for Experimental Software Engineering (IESE), Kaiserslautern. Since November 2001, Dr. Münch has been an executive board member of the temporary research institute SFB 501 "Development of Large Systems with Generic Methods" funded by the German Research Foundation (DFG). Dr. Münch’s research interests in software engineering include: (1) modeling and measurement of software processes and resulting products, (2) software quality assurance and control, (3) technology evaluation through experimental means and simulation, (4) generic methods for the development of large systems, (5) technology transfer methods. He has been teaching and training in both university and industry environments, and also has significant R&D project management experience. Jürgen Münch is a member of the IEEE Computer Society and the German Computer Society (GI).

**A. Ocampo** Alexis Ocampo received his MSc degree in Computer Science from Los Andes University, Colombia, in 1999 and his title as Systems Engineer from Industrial University of Santander, Colombia, in 1997. Since February 2002, Alexis Ocampo has been a research scientist at the Fraunhofer Institute for Experimental Software Engineering (IESE), Kaiserslautern, in the department of Quality and Process Engineering. Before that, he worked for 5 years as a research-developer on new technologies and methodologies with the software company Heinsohn Associates, Bogota, Colombia. His master thesis entitled “Implementation of PSP in the Colombian Industry: A case study” was developed within this company. He also worked as an instructor at the University of Los Andes in the Department of Systems and Computation. Alexis Ocampo’s research interests in software engineering include: (1) modeling and measurement of software processes and resulting products, (2) software quality assurance and control, (3) technology transfer methods.